\documentstyle[11pt,Mor96]{article}
\def\ltsima{$\; \buildrel < \over \sim \;$}
\def\simlt{\lower.5ex\hbox{\ltsima}}
\def\gtsima{$\; \buildrel > \over \sim \;$}
\def\simgt{\lower.5ex\hbox{\gtsima}}
\def\d{\hbox{d}}
\def\ii{\hbox{i}}

\def\vu{\hbox{\bf u}}
\def\vx{\hbox{\bf x}}
\def\vk{\hbox{\bf k}}

\def\d{\hbox{d}}

\def\mg{\big<}
\def\md{\big>}

\def\gm{\gamma}
\begin{document}
\heading{Large-Scale Structure Formation in the Quasi-linear Regime} 

\author{F. Bernardeau}
       {Service de Physique Th\'eorique, C.E. de Saclay, }
       {F-91191 Gif-sur-Yvette, France}

\begin{abstract}{\baselineskip 0.4cm 
The understanding of the large-scale structure formation requires the
resolution of coupled nonlinear equations describing the cosmic density
and velocity fields. This is a complicated problem that,
for the last decade, has been essentially addressed with N-body simulations.
There is however a regime, the so-called quasi-linear regime,
for which the relative density fluctuations are on average
below unity. It is then possible to apply Perturbation Theory techniques
where the perturbation expansions are made with respect to the initial
fluctuations. I review here the major results that have been obtained 
in this regime.
}

\end{abstract}

\section{The Gravitational Instability Scenarios}

Perturbation Theory (PT) has had important developments in the last 
few years simultaneously from theoretical, numerical and
observational points of view. These techniques allow indeed
to explore nonlinear features revealed in the statistical properties of the 
large-scale structures of the Universe, for which 
precise and robust data are now available.
The general frame of these calculations is based on the dynamics of a 
self-gravitating {\sl pressure-less} fluid. The large-scale 
structures are then assumed to have gravitationally
grown from small initial fluctuations.
 It is important to note that in the following 
these initial fluctuations will be assumed
to follow a Gaussian statistics. That excludes a priori exotic
models that make intervene topological defects as seeds of
structure formation.

By linearizing the field equations, i.e. when the quadratic
coupling between the fields is neglected, one can compute
the growth rate of the fluctuations.
This allows to compare for instance the
Cosmic Microwave Background temperature anisotropies as measured
by the COBE satellite with the local density fluctuations
observed in galaxy catalogues. The linear  approximation cannot be
used however to explore the statistical properties of the density field:
the local fluctuations are just amplified, their shapes
are not changed, and therefore the linear 
density field remains Gaussian if it obeyed such statistics
initially.

\section{The Perturbation Theory}

The principles of these calculations have been initially
presented by Peebles \cite{Peeb}, then explored in more details
in a series of recent papers \cite{Fry84, Gor86, Bou92, Ber92, 
JBC, Jusz, Mak, Ber94, Scocci1, Scocci2}.
The starting point of all these calculations is the system
of field equations, Continuity, Euler and Poisson equations, 
describing a single stream pressure-less fluid. 

The density and velocity fields which satisfy 
those equations are then expanded with respect
to the initial fluctuation field,
\begin{equation}
\delta(t,\vx)=\sum_i\delta^{(i)}(t,\vx),\ \ \ \ 
\vu(t,\vx)=\sum_i\vu^{(i)}(t,\vx).\label{dev}
\end{equation}
The fields $\delta^{(1)}(t,\vx)$ and $\vu^{(1)}(t,\vx)$
are just the local linearized density and velocity fields.
They are linear in the initial density field.
The higher order terms, $\delta^{(2)}$, $\delta^{(3)}$,.. are
respectively quadratic, cubic,.. in the initial density
field, and, therefore, do not obey a Gaussian 
statistics.

\subsection{The Density Field}

The time and space dependences of 
the linearized density field, $\delta^{(1)}$,
factorize so that it can be written,
\begin{equation}
\delta^{(1)}(t,\vx)=D(t)\,\int\d^3\vk
\,\delta(\vk)\,\exp(\ii\vk\vx),
\end{equation}
where $\delta(\vk)$ are the Fourier transforms of the initial
density field. They are assumed to form a set of
Gaussian variables. Their statistical properties are then entirely
determined by the shape of the power spectrum, $P(k)$,
defined  by,
\begin{equation}
\mg\delta(\vk)\,\delta(\vk')\md=\delta_{\rm Dirac}(\vk+\vk')\,P(k),
\end{equation}
where $\mg.\md$ denotes ensemble averages over the initial conditions.
The function $D(t)$ is determined by the cosmological parameters.
It is proportional to the expansion factor $a(t)$ for an Einstein-de Sitter
universe. In general it has been found to depend on the 
cosmological density $\Omega$ in such a way that
$(\d\log D/\d\log a)\approx\Omega^{0.6}$ \cite{Peeb} and to be
very weakly dependent on the cosmological constant $\Lambda$ \cite{Lah}.

The higher order terms can all be recursively obtained from the
linear term. In general one can write the $i^{\rm th}$ order term
as \cite{Gor86},
\begin{equation}
\delta^{(i)}(t,\vx)=
D^i(t)\int\d^3\vk_1\dots\d^3\vk_i
\,\delta(\vk_1)\dots\delta(\vk_i)\,\exp[\ii\vx\cdot(\vk_1+\dots+\vk_i)]\,
F(\vk_1,\dots,\vk_i),\label{deltai}
\end{equation}
where $F(\vk_1,\dots,\vk_i)$ is a homogeneous function of
the angles between the different wave vectors. Note that
the time dependence given here is only approximate for 
a non Einstein-de Sitter Universe. 
It is however a very good approximation \cite{Ber92,Bou92,Ber94a,Ber94}.

\subsection{The Velocity Field}

For the velocity field the situation is very similar.
We have first to notice that in the single flow approximation
the vorticity is expected to be diluted by the expansion (eg. \cite{Peeb})
and thus to be negligible {\sl at any order} of the perturbation 
expansion. Then it is more natural to present the properties of
the velocity field in terms of the local divergence (expressed in units
of the Hubble constant), 
\begin{equation}
\theta(t,\vx)\equiv{\nabla_{\vx}\cdot\vu(t,\vx)\over H(t)}.
\end{equation}
We then have,
\begin{equation}
\theta^{(1)}(t,\vx)={\d\log D\over\d\log a}\ \delta^{(1)}(t,\vx)\approx
\Omega^{0.6}(t)\ \delta^{(1)}(t,\vx).\label{vdensrel}
\end{equation}
The higher order terms can be written
\begin{equation}
\theta^{(i)}(t,\vx)\approx\Omega^{0.6}(t)\ D^i(t)
\int\d^3\vk_1\dots\d^3\vk_i
\,\delta(\vk_1)\dots\delta(\vk_i)\,\exp[\ii\vx\cdot(\vk_1+\dots+\vk_i)]\,
G(\vk_1,\dots,\vk_i),\label{thetai}
\end{equation}
where $G$ is another homogeneous function, different from $F$.
Note that in general the time dependence of
$\theta^{(i)}$ is not the power $i$ of the time dependence 
of $\theta^{(1)}$. We will see that it induces 
remarkable statistical properties for the local 
velocity field.

\subsection{Implications for the Statistical Properties of the Cosmic Fields}

In general the consequences of the existence
of higher order terms can be separated in two categories:
these terms affect the mean growth rate
of the fluctuations and introduce new statistical properties
because of their non-Gaussian nature. I will briefly review
both aspects here.

\section{The Mean Growth Rate of the Fluctuations}

The mean growth rate of the fluctuation can be calculated from the 
shape and magnitude of the variance of the local density
contrast \cite{J81,Lokas,Scocci2}, or equivalently 
of the evolved power spectrum \cite{Mak,Jain}. For instance the variance, 
$\mg\delta^2\md$, can be calculated using the expansion in (\ref{dev}),
\begin{equation}
\mg\delta^2\md=\mg\left(\delta^{(1)}+\delta^{(2)}+\dots\right)^2\md.
\end{equation}
By re-expanding this expression with respect to the initial density
field one gets,
\begin{equation}
\mg\delta^2\md=\mg\left(\delta^{(1)}\right)^2\md+\ 2\ 
\mg\delta^{(1)}\ \delta^{(2)}\md+
\mg\left(\delta^{(2)}\right)^2\md+\ 2\ 
\mg\left(\delta^{(1)}\ \delta^{(3)}\right)^2\md+\dots
\end{equation}
The term $\mg\delta^{(1)}\ \delta^{(2)}\md$ is zero
because it is cubic in the initial Gaussian variables. 
The calculation of the corrective terms 
(the so-called ``loop corrections'', see \cite{Gor86,Ber92,Scocci1,Scocci2}),
$\mg\left(\delta^{(2)}\right)^2\md+2
\mg\delta^{(1)}\ \delta^{(3)}\md$,
can be done from the recursive solution of the field equations.
In practice these calculations become rapidly very complicated
but they can be simplified in case of a power law spectrum,
\begin{equation}
P(k)\propto\ k^n.
\end{equation}

Thus for $n=-2$ one gets
\begin{equation}
\mg\delta^2\md=\mg\delta_{\rm lin}^2\md+{1375\over 1568}\ 
\mg\delta_{\rm lin}^2\md^2\approx\mg\delta_{\rm lin}^2\md+0.88\ 
\mg\delta_{\rm lin}^2\md^2
\label{lcorr}
\end{equation}
where $\delta_{\rm lin}$ is the linearized density contrast, 
$\delta_{\rm lin}=\delta^{(1)}$.
This prediction can be compared with numerical results obtained from $N$-body
codes. An interesting way to compared the two is to use the
phenomenological description of the nonlinear growth rate of the
fluctuations proposed by Hamilton et al. \cite{hamil}.
This description relates the mean linear density contrast at the
scale $R_{\rm lin}$ to the full non-linear one at the scale
$R_{\rm non-lin}$ defined by
\begin{equation}
R_{\rm non-lin}^3\ \left(1+\mg\delta^2\md\right)
=R_{\rm lin}^3.
\end{equation}
The universal transform proposed by Hamilton et al. 
(given here in terms of the power spectrum)
is presented in Fig. 1 (solid line) and compared to the
prediction (\ref{lcorr}), (short dashed line). 
The latter gives remarkably well 
the position of the departure from the pure linear regime (straight line).
On the other hand the position of this
transition is quite poorly given by the Zel'dovich approximation
(long dashed line).

\medskip
\vglue 11 truecm
\special{hscale=70 vscale=70 voffset=-60 hoffset=45 psfile=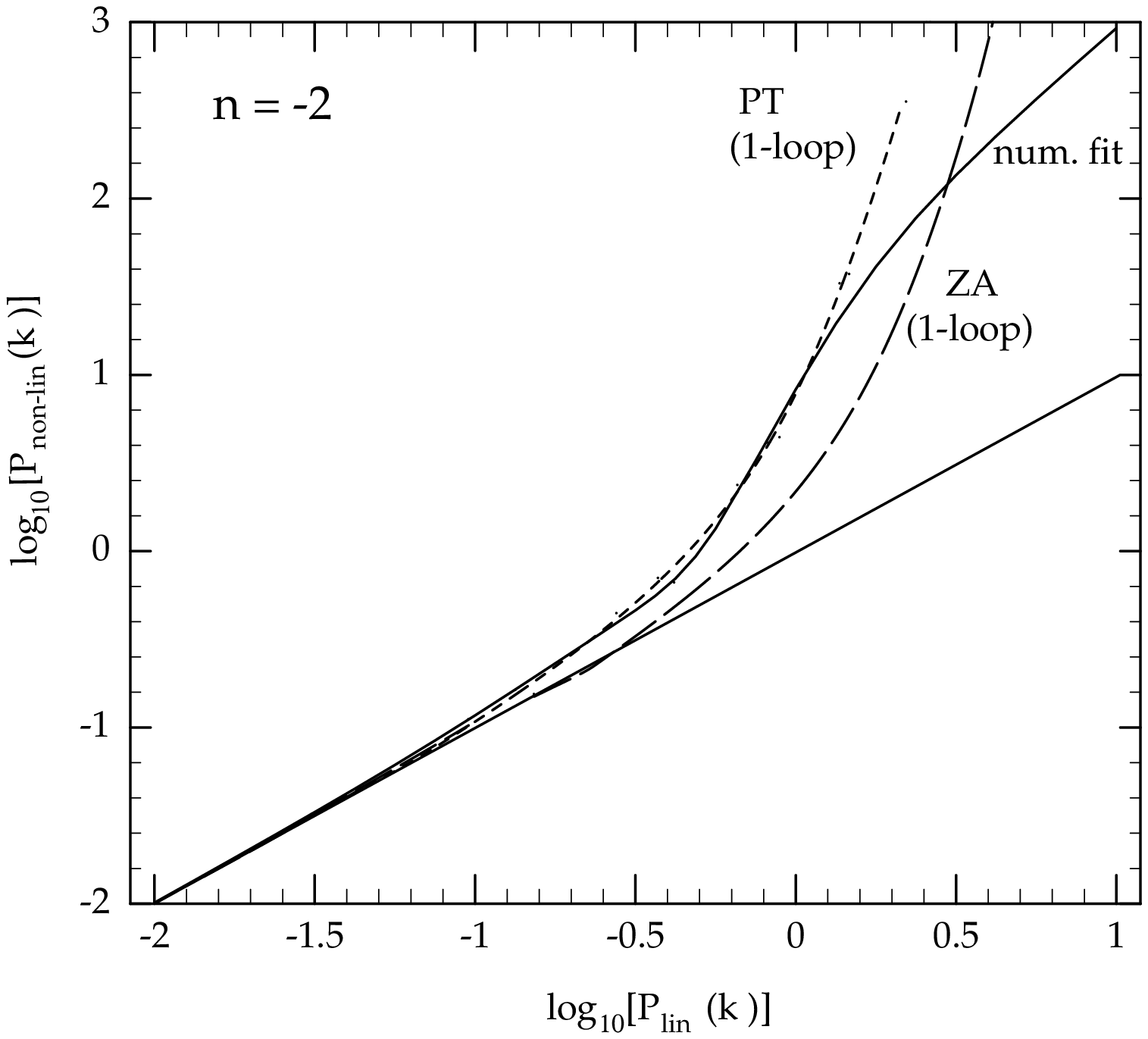}
\begin{center}
\narrower
{\bf Figure 1.} {\baselineskip 0.4cm 
\textwidth 15 truecm
Comparison of the PT predictions (\ref{lcorr})
with the Hamilton et al. \cite{hamil} prescription
for the growth rate of the fluctuation (solid line) in case of
$P(k)\propto k^{-2}$. The short dashed line
is the prediction of the next-to-leading order Perturbation Theory 
result \cite{Scocci2}},
and the straight line is the linear theory prediction
(figure taken from \cite{Scocci2}).
\end{center}
\medskip

Surprisingly the corrective terms are finite only for
$n<-1$. For $n$ larger than $-1$ the loop terms contain a divergence 
so that the result explicitly depends on a cutoff $k_c$ introduced
in the shape of the power spectrum at large $k$. Thus 
when $(R\ k_c)$ is large we have,
\begin{equation}
\mg\delta^2\md=\mg\delta_{\rm lin}^2\md+C_n\ \left( k_c\,R\right)^{n+1}\ 
\mg\delta_{\rm lin}^2\md^2
\end{equation}
where $R$ is the filtering radius.
The existence of this divergence was not really expected since
the numerical results do not indicate any significant change of
behavior for $n\approx -1$. One could argue that in practice this
is not a relevant problem since in realistic scenarios
(like the CDM power spectrum) the index reaches
$n=-3$ at small scale, so that the corrective terms are naturally
regularized. However with such an interpretation it implies 
that the growth of the very large scale fluctuations is fed
with smaller scale fluctuations, which would
be slightly surprising in view of the numerical results
accumulated over the last few years. Another possible explanation is that
the higher order corrections induce other divergences that
cancel each other (a phenomenon quite common in statistical physics).
Then the corrective terms to the linear growth rate  
would not be proportional to  the square of
$\mg\delta_{\rm lin}^2\md$ but to a smaller power of it.
Future analytic investigations may be able to throw light on this
problem.

\section{The Emergence of non-Gaussian Features}

\subsection{The Moments}

The other major consequence of the existence of non-linear terms
in (\ref{dev}) is the apparition of non-Gaussian properties.
Although it is possible to characterize non-Gaussian features
in many different ways, most of the efforts have been devoted
to properties of the one-point
probability distribution function (PDF) of the local density. 
More particularly, attention has been focussed on the moments
if this distribution and how they are sensitive to
non-linear corrections.
In his treatise \cite{Peeb}, Peebles already considered
the implications of second order perturbation theory for the
behavior of the third moment of the local density in case
of initial Gaussian fluctuations.
Indeed, using the expansion (\ref{dev}), the third moment reads,
\begin{equation}
\mg\delta^3\md=\mg\left(\delta^{(1)}+\delta^{(2)}+\dots\right)^3\md
=\mg\left(\delta^{(1)}\right)^3\md
+\ 3\ \mg\left(\delta^{(1)}\right)^2\delta^{(2)}\md+\dots,
\end{equation}
and what should be the dominant term of this series,
$\mg\left(\delta^{(1)}\right)^3\md$, is identically zero
in case of Gaussian initial conditions. As a consequence
the third moment is actually given by
\begin{equation}
\mg\delta^3\md\approx\ 3\ \mg\left(\delta^{(1)}\right)^2\delta^{(2)}\md.
\end{equation}
Peebles calculated this expression neglecting the effects of smoothing
and found (for an Einstein-de Sitter Universe),
\begin{equation}
\mg\delta^3\md\approx\ {34\over 7}\ \mg\delta_{\rm lin}^2\md^2,
\end{equation}
so that the ratio, $\mg\delta^3\md/\mg\delta^2\md^2$,
is expected to be finite at large scale. 
It is actually possible to extend this
quantitative behavior to higher order moment and to show that 
the cumulants (i.e., the connected parts of the moments)
are related to the second moment so that the ratios,
\begin{equation}
S_p=\mg\delta^p\md_c/\mg\delta^2\md^{p-1},
\end{equation}
are all finite at large scale. As mentioned before,
the coefficient $S_3$ was computed by
Peebles \cite{Peeb}, Fry \cite{Fry84} derived $S_4$ and
eventually Bernardeau \cite{Ber92} gave the whole series
of these coefficients.

Unfortunately these early calculations did not take into account
the filtering effects, that is that the ensemble averages should be done on
the local smoothed fields.
This problem was addressed numerically
by Goroff et al. \cite{Gor86} for a Gaussian window function for the
third and fourth moments. More recently these two coefficients
have been calculated analytically and semi-analytically in
\cite{Jusz,Lok} for this window function. However, the results
turn out to be simpler in case of a top-hat window, as it was noticed
by Juszkiewicz et al. \cite{JBC} for $S_3$ and power law spectra.
The coefficients $S_3$ and $S_4$ were calculated for this window
in \cite{Ber94a} for any spectrum and any cosmological
models. Eventually Bernardeau \cite{Ber94}
proposed a method to derive the whole series of these
coefficients from the spherical collapse dynamics.

I recall here the expression of the first two coefficients $S_3$ and $S_4$,
as a function of the shape of the second moment,
\begin{equation}
S_3(R)={{34}\over 7} + {\gm}(R);
\label{s3}
\end{equation}
\begin{equation}
S_4(R)= {{60712}\over {1323}} + {{62\,{\gm(R)}}\over 3} + 
   {{7\,{{{\gm(R)}}^2}}\over 3} + {2\,\over 3}
{\d\gm(R)\over\d\log R};
\label{s4}
\end{equation}
with 
\begin{equation}
\gm={\d\log\sigma^2(R)\over\d\log R}.\label{eq:gam}
\end{equation}
One can notice that $S_3$ depends only on the local
slope, and that $S_4$ also depends (but weakly) on the variations
of that slope. These results are valid for an Einstein-de Sitter
Universe but are found to be weakly $\Omega$ and $\Lambda$ dependent
\cite{Bou92,Ber92,Ber94a}.

\medskip
\vglue 14 truecm
\special{hscale=80 vscale=80 voffset=0 hoffset=-10 psfile=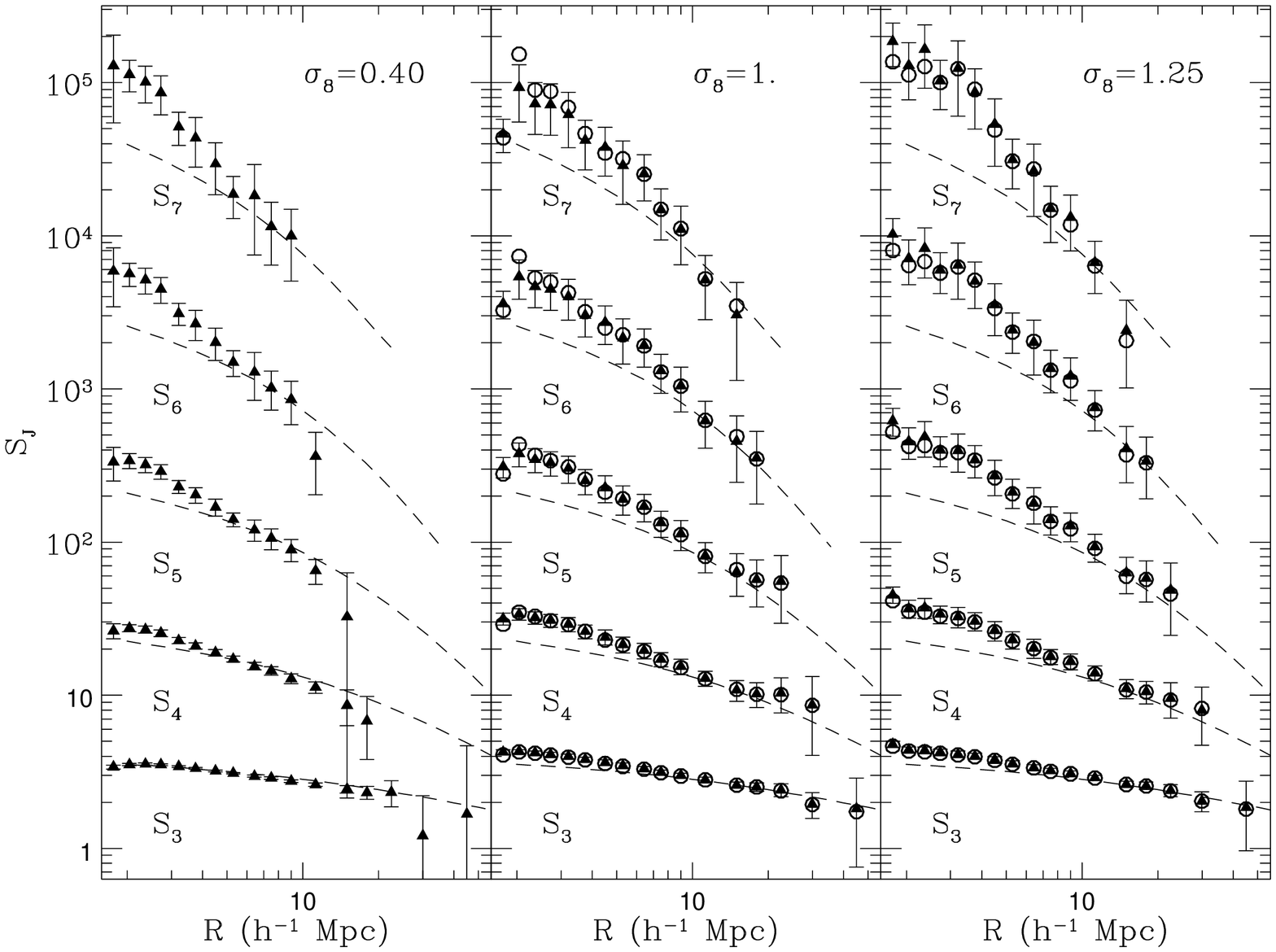}
\begin{center}
\narrower
{\bf Figure 2.} {\baselineskip 0.4cm 
\textwidth 15 cm
Comparison of PT predictions
with results of numerical simulations with CDM
initial conditions for moments  of the local density contrast. 
The PT results (dashed lines) are compared to the
measured $S_p$ coefficients as a function of radius for
three different time-steps (figure taken 
from \cite{Baugh}). 
}
\end{center}
\medskip

These coefficients have been tested against numerical results 
by different groups \cite{Bou92,Ber94a,Baugh,Lok} and have
been found to be in extremely good agreement with the numerical 
results. In particular as shown in Fig.2, 
it has been found that the measured $S_p$
coefficients are in good agreement with the PT results
up to $p=7$ as long as the variance $\mg\delta^2\md^{1/2}$
is lower than unity \cite{Baugh}. This constraint does not seem to become
tighter for higher orders.

\subsection{From Cumulants to PDF-s}

The shape of the PDF and the values of the cumulants are obviously
related. When a limited number of cumulants is known it is possible
to reconstruct some aspect of the shape of the PDF using the so-called
Edgeworth expansion (see \cite{Jusz,BerKof}). But actually as the
whole series of the cumulants is known for top-hat filtering
it is possible
to invert the problem (at least numerically) and build the
PDF from the generating function of the cumulants (see \cite{BalS}
for the general method and \cite{Ber92,Ber94} for the
application to PT results). This method is justified by the fact that
the $S_p$ coefficients converge to their asymptotic PT values
at roughly the same rate, i.e. for the same values of $\mg\delta^2\md$
(Fig. 2).

\medskip
\vglue 10 truecm
\special{hscale=95 vscale=95 voffset=-13 hoffset=0 psfile=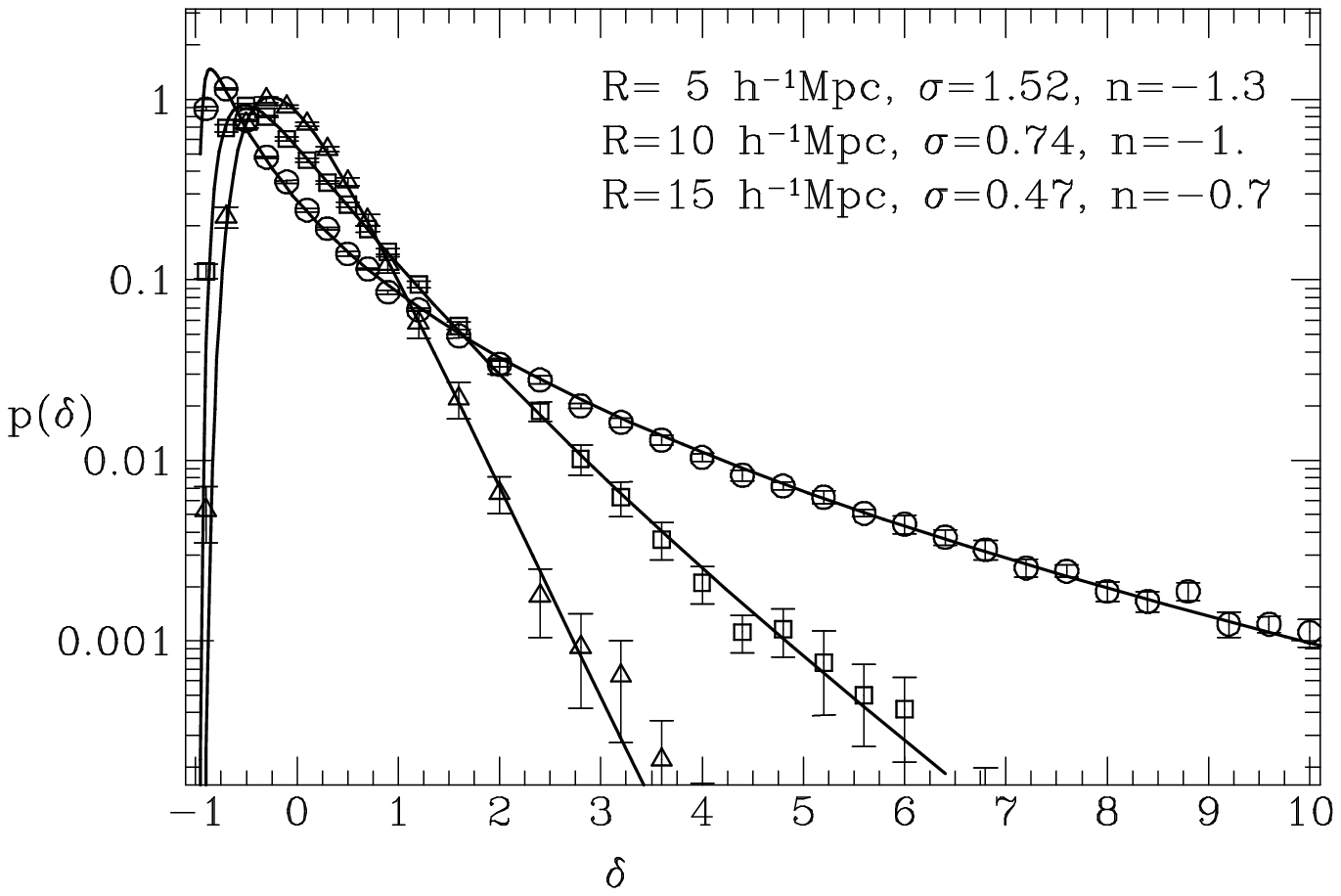}
\begin{center}
\narrower
{\bf Figure 3.} {\baselineskip 0.4cm 
\textwidth 15 cm
Comparison of PT predictions
with results of numerical simulations with CDM
initial conditions for the shape of 
the one--point PDF of the density contrast. 
The density PDF has been measured at three different radius, $R=5$ 
h$^{-1}$Mpc (circles),
$R=10$ h$^{-1}$Mpc (squares), $R=15$ h$^{-1}$Mpc (triangles)
and compared to the PT predictions calculated from the
measured values of the variance and the local index
(figure taken from \cite{Ber94}).
}
\end{center}
\medskip

In Fig. 3, I present a comparison of the PT results obtained
in such a way to the numerical measurements of PDF-s. Three
different smoothing radii have been chosen. For each radius
one can get the value of the initial index, and the value of the variance.
The predicted PDF-s are entirely determined by those two numbers.
As it can be seen the agreement is extremely good.

\section{Comparison with data}

These successes have given confidence in the validity of the PT results
and have boosted investigations in the available
galaxy catalogues.
Using the IRAS galaxy catalogue Bouchet et al. \cite{Bou}
measured the $3^{\rm rd}$ and $4^{\rm th}$ moment of the one-point PDF of the
local galaxy density. 
The observed relation between the
third moment and the second is the following,
\begin{equation}
\mg\delta_{\rm IRAS}^3\md\ \approx\ 1.5\ \mg\delta_{\rm IRAS}^2\md^2.
\end{equation}
It is a strong indication in favor of Gaussian initial 
conditions since the exponent is indeed the one expected from
PT. The coefficient however is lower from what is expected, but
the comparison with quantitative predictions is 
complicated for two reasons. The galaxy positions are known only
in redshift space (that is that their distance is assumed to be
proportional to their line-of-sight velocity), and IRAS galaxies might be
strongly biased with respect to the mass distribution.
The first problem has been addressed in \cite{Hivon}
where it is shown with adapted PT
calculations that the effects of redshift space distortion on $S_3$ are
small at large scale. It has been also investigated
numerically in \cite{Lahav}. 
The fact that the measured coefficient, 1.5, is smaller
than what is expected (by a factor of about 1.7) is then
probably due to biases in the galaxy distribution. 
In case of the IRAS galaxies this is not too surprising since
those galaxies are known to be under-populated (compared to bright galaxies)
in very dense areas.

For these reasons a lot of interest has been devoted recently
to the APM angular galaxy catalogue. The fact that it has more than 
1,300,000 objects makes it the largest galaxy catalogue now available
and a perfect domain of investigation. Measurements of the
$S_p$ coefficients have been made by Gazta\~naga \cite{gaz} for $p\le7$.
Comparison with PT predictions are however not straightforward
because of the projections effects with which the relations 
(\ref{s3},\ref{s4}) for $S_3$ and $S_4$ are not valid anymore.

Adapted calculations that take into account this new geometry and
using the small angle approximation have been made in \cite{Ber95}
for any order of cumulants but assuming a power law spectrum.
These results have been recently extended in \cite{Pollo}
for any shape of power spectrum. In fig 4. I present the comparison
of the measured $S_3$ and $S_4$ coefficients as a function of the 
smoothing angle (triangles) compared to the predicted ones from PT
(solid lines). The latter have been calculated with either the
small angle approximation or with a direct Monte-Carlo integration
for which no such approximation is required (for $S_3$ only).

\medskip
\vglue 7.5 truecm
\special{hscale=70 vscale=70 voffset=-122 hoffset=-20 psfile=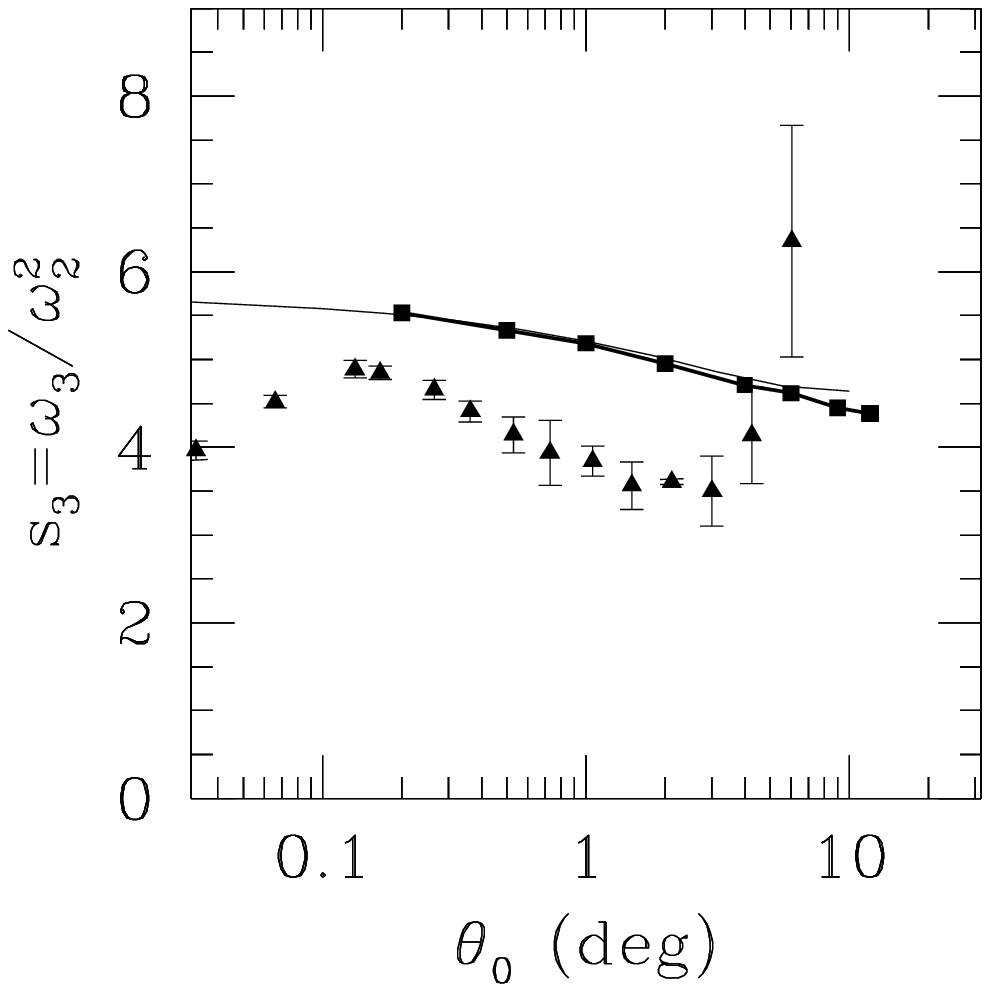}
\special{hscale=70 vscale=70 voffset=-122 hoffset=230 psfile=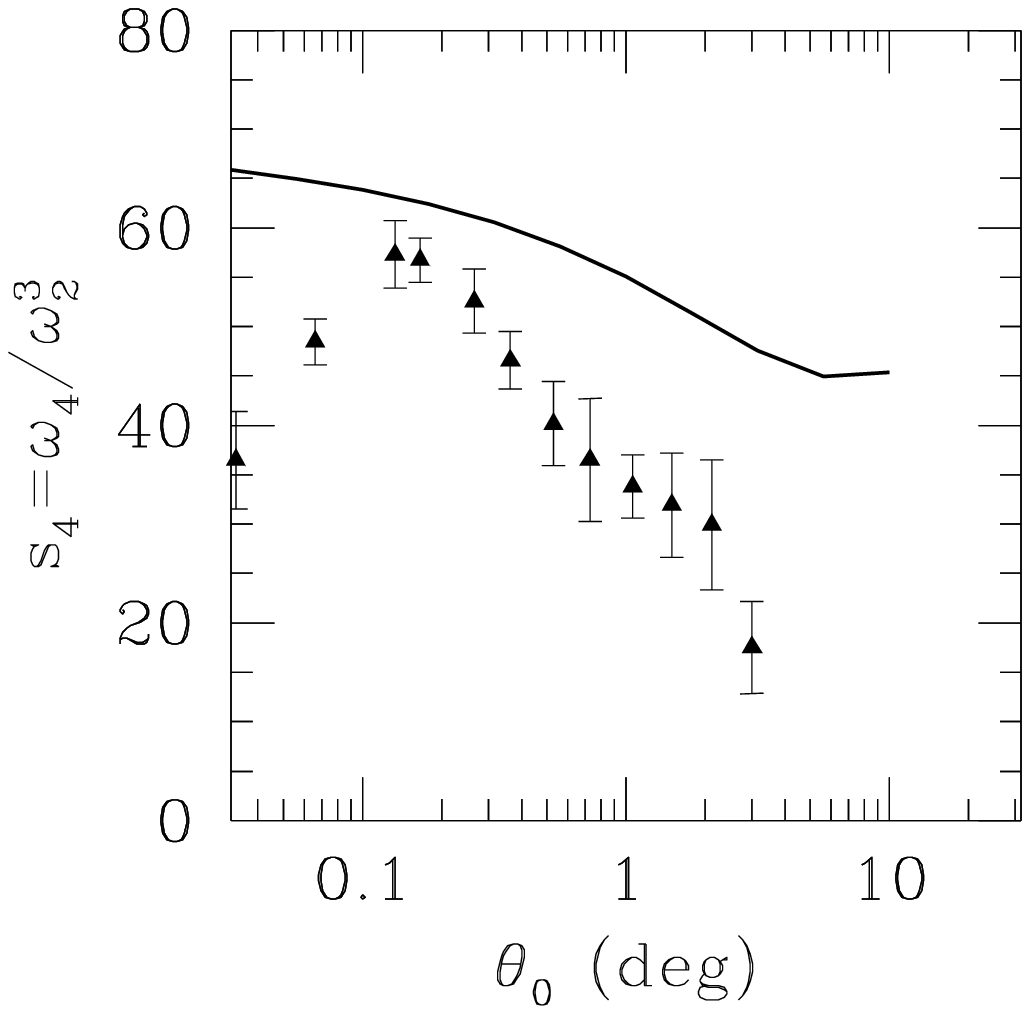}
\begin{center}
\narrower
{\bf Figure 4.} {\baselineskip 0.4cm 
\textwidth 15 cm
The $s3(\theta)$ and $s4(\theta)$ coefficients in the APM angular catalogue
(triangles, see \cite{gaz}) compared to the PT predictions, with
the small angle approximation (solid lines), or with a numerical 
integration of the projection effects (squares).}
\end{center}
\medskip

There is still a significant discrepancy between the PT results
and the observations. As the theoretical predictions are
very robust with respect to uncertainties in the selection function
or with respect to values of $\Omega$ and $\Lambda$ (see
\cite{Ber95}), 
it does not seem possible
to reconcile the PT predictions with the observations, unless
galaxies are biased. This bias is however less important than the 
one expected for the IRAS galaxies. 
These observations are anyway a strong indication in favor
of Gaussian initial conditions. The implications of models
with topological defects like string theories or textures 
have been investigated in particular by Gazta\~naga \cite{gaz96}.
He found that the current data, unless the galactic distribution
is unrelated to the mass distribution(!), exclude such theories.

\section{The Velocity Field Statistics}

Although data on the peculiar velocities are more difficult to obtain, 
it is a very interesting domain of investigation since 
cosmic velocities are in principle directly related to the mass fluctuations.
In a recent review, Dekel \cite{dek94} has presented the observational
and theoretical status of this rapidly evolving field. 

The line-of-sight peculiar velocities are estimated from the Tully-Fisher
(or similar) relation between the absolute luminosity of the galaxies and
their internal velocity dispersion for elliptical, or their
circular velocity for spirals. From these informations, and taking advantage 
of the expected absence of vorticity, it is possible to build the whole 3D
smoothed velocity field. It has been done in particular
in \cite{pot1,pot2}, using the so-called POTENT reconstruction method.
A straightforward application of the reconstructed velocity field
is the use of the velocity-density relationship obtained
in the linear regime (\ref{vdensrel}) that would indeed provide a way to 
measure $\Omega$. This is possible however only if galaxies
are not biased. Otherwise, assuming  that at large scale,
$\delta_{\rm galaxies}^{(1)}(\vx)=b\ \delta^{(1)}(\vx)$,
the observations  constrain a combination of $\Omega$ and $b$,
$\beta={\Omega^{0.6}/b}$.
Various results for $\beta$ have been given in the literature (see
\cite{dek94}). A rough compilation of them leads to $0.3\lsim\beta\lsim 1.2$.

As we have no robust models for the large-scale bias of the
galaxies, it is quite natural to explore the {\sl intrinsic} properties
of the velocity field. In the following, no assumptions are
made on the bias, galaxies are simply assumed to act as {\em test particles}
for the large-scale flows.
Within this scheme Dekel \& Rees \cite{DekRees}
proposed to use the maximum expanding void to put constraints on $\Omega$;
Nusser \& Dekel \cite{Nusser}
tried to reconstruct the initial density field using the Zel'dovich
approximation, thus constraining $\Omega$ on the basis of
Gaussian initial conditions. Here I present a more systematic
study of the expected properties of the local divergence using PT. 

In a similar way than for the density field it is indeed possible
to compute the coefficients $T_p$ that relate the high order moments of the
local divergence to the second moment,
\begin{equation}
T_p=\mg\theta^p\md_c/\mg\theta^2\md^{p-1}.
\end{equation}
Unlike the $S_p$
coefficients these ones are found to be strongly $\Omega$ dependent
(but weakly $\Lambda$ dependent),
\begin{equation}
T_p(\Omega)\propto\Omega^{-0.6(p-2)},
\end{equation}
as is a direct consequence of the time dependence of $\theta^{(i)}$
found in (\ref{thetai}). The parameter $T_3(\Omega)$ was calculated
in \cite{vskew} as a function of $\Omega$,
\begin{equation}
T_3(\Omega)={1\over \Omega^{0.6}}\left[{26\over 7}
+(n+3)\right],
\end{equation}
and proposed as a possible indicator to measure $\Omega$.
A preliminary investigation using the POTENT data gave $\Omega>0.3$
with a good confidence level.

\medskip
\vglue 7.5 truecm
\special{hscale=88 vscale=88 voffset=-270 hoffset=0 psfile=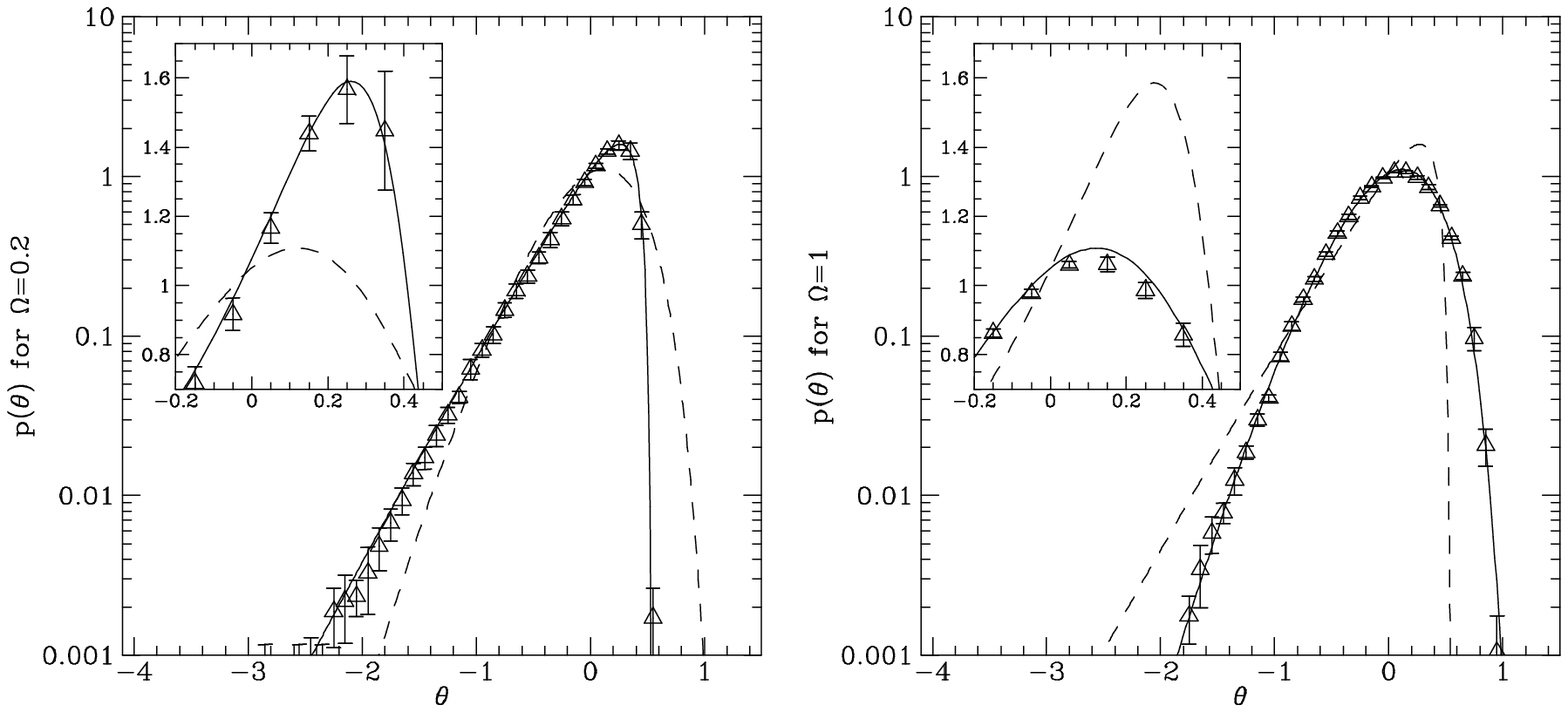}
\begin{center}
\narrower
{\bf Figure 5.} {\baselineskip 0.4cm 
\textwidth 15 cm
Comparison of the PT predictions for the shape
of the PDF of the velocity divergence (solid lines)
with results of numerical simulations for $n=-1$. 
In the left panel $\Omega=0.2$ and in the right $\Omega=1$. In both
cases we have $\sigma_{\theta}\approx0.4$. The dashed lines are
the PT predictions when the reverse assumption is made on $\Omega$.
}
\end{center}
\medskip

But more generally 
the $\Omega$ dependence of the $T_p$ coefficients
obviously extends to the shape of the PDF
of the local divergence. In particular we can show that for a
power law spectrum with $n\approx -1$, it is approximately given by,
\begin{eqnarray}
\lambda&=&1-{2 \theta\over 3 \Omega^{0.6}},\ \ 
\kappa=1+{\theta^2\over 9 \lambda \Omega^{1.2}};\\
p(\theta)&=&{\left(
[2 \kappa-1]/\kappa^{1/2}+[\lambda-1]/\lambda^{1/2}\right)^{-3/2}\over
\kappa^{3/4} (2 \pi)^{1/2}\,\sigma}\ 
\exp\left[-{\theta^2\over 2 \lambda \sigma^2}\right],
\end{eqnarray}
where $\sigma$ is the rms of the fluctuations of $\theta$. This
distribution exhibits a strong $\Omega$ dependence as it can be seen
on Fig. 5. As expected, when $\Omega$ is low the distribution is strongly
non-Gaussian, with a sharp cut-off in the large
divergence tails (it corresponds to rapidly expanding voids). 
This is this feature
(discovered in an independent way) that was used in \cite{DekRees}
to constrain $\Omega$.

We have checked these theoretical predictions with numerical
experiments (see Fig. 5) and found once again that 
the numerical results  are in excellent agreement with them.
It opens ways to have reliable measure of $\Omega$ 
from velocity data.

\section{Conclusions}

As this rapid overview have shown it, the study of the quasi-linear
regime is in rapid development. A lot of efforts have been devoted to 
analytic calculations, and our understanding of the
growth of structures in the intermediate regime has considerably
improved. In particular comparisons with numerical simulations have
shown that the PT results have a surprisingly large validity domain
for the density as well as for the velocity fields.
So far, most of the comparisons have been done for the cumulants at their 
leading order. However, the recent analytic results obtained 
for the next-to-leading order have open puzzling questions for 
their interpretation.

In any case, these PT results provide extremely discriminatory tools to test 
the gravitational instability scenarios. Comparisons with data 
have already provided valuable insights into the properties of the galaxy
distribution: they have indeed given strong indications 
in favor of Gaussian initial
conditions and have pointed out precious indications on the existence
and nature of biases between the large-scale matter
distribution and the galaxy distribution.
Moreover, assuming Gaussian initial conditions, it seems possible 
to get reliable {\em and bias independent} 
constraints on $\Omega$ from the statistics 
of the large-scale cosmic flows.

\acknowledgements{I would like to thank Roman Scoccimarro and
Enrique Gazta\~naga for permission to include some of their figures.}

\resume{
La compr\'ehension de la formation des grandes structures
requiert la r\'esolution d'\'equations non-lin\'eaires coupl\'ees
d\'ecrivant l'\'evolution des champs de densit\'e et de vitesse
cosmologiques. C'est un probl\`eme compliqu\'e qui, ces dix derni\`eres
ann\'ees, a \'et\'e trait\'e essentiellement avec des simulations
num\'eriques \`a $N$ corps. Il y a cependant un r\'egime, le r\'egime
dit quasi-lin\'eaire, pour lequel les fluctuations relatives de densit\'e
sont inf\'erieures \`a l'unit\'e en moyenne. Il est alors possible
d'utiliser des techniques de th\'eorie des perturbations o\`u
les d\'eveloppements perturbatifs sont faits par rapport
aux fluctuations initiales. Je pr\'esente ici les r\'esultats
majeurs qui ont \'et\'e obtenus dans ce r\'egime.
}

\vfill
\end{document}